\documentclass[aip, apl, amsmath, amssymb, reprint]{revtex4-1}

\usepackage{graphicx}
\usepackage{dcolumn}
\usepackage{bm}

\usepackage[utf8]{inputenc}
\usepackage[T1]{fontenc}
\usepackage{mathptmx}
\usepackage{etoolbox}
\usepackage{xcolor}
\usepackage{subfig}
\usepackage{ragged2e}

\makeatletter
\def\@email#1#2{%
 \endgroup
 \patchcmd{\titleblock@produce}
  {\frontmatter@RRAPformat}
  {\frontmatter@RRAPformat{\produce@RRAP{*#1\href{mailto:#2}{#2}}}\frontmatter@RRAPformat}
  {}{}
}%
\makeatother
\begin{document}

\preprint{AIP/123-QED}

\title[MgB$_2$ TKID]{MgB$_2$ Thermal Kinetic Inductance Detector}
\author{T. Jabbari}
\thanks{T. Jabbari and A. Hawkins contributed equally to this work.}
\affiliation{ Jet Propulsion Laboratory, California Institute of Technology, Pasadena, CA 91109, USA}

\author{A. Hawkins}
\thanks{T. Jabbari and A. Hawkins contributed equally to this work.}
\affiliation{%
Department of Physics, Arizona State University, Tempe, AZ 85287, USA
}

\author{A. Wandui}%
\affiliation{ Jet Propulsion Laboratory, California Institute of Technology, Pasadena, CA 91109, USA}
 
\author{C. Frez}%
\affiliation{ Jet Propulsion Laboratory, California Institute of Technology, Pasadena, CA 91109, USA}

\author{J. Greenfield}
\affiliation{ Jet Propulsion Laboratory, California Institute of Technology, Pasadena, CA 91109, USA}
\affiliation{School of Earth and Space Exploration, Arizona State University, Tempe, Arizona 85287, USA   }

\author{C. Roberson}
\affiliation{School of Earth and Space Exploration, Arizona State University, Tempe, Arizona 85287, USA   }

\author{M.J. Lee}
\affiliation{School of Earth and Space Exploration, Arizona State University, Tempe, Arizona 85287, USA   }

\author{P. Mauskopf}
\affiliation{School of Earth and Space Exploration, Arizona State University, Tempe, Arizona 85287, USA   }
\affiliation{%
Department of Physics, Arizona State University, Tempe, AZ 85287, USA
}

\author{D. Cunnane}
\thanks{Author to whom correspondence should be addressed.}
\affiliation{ Jet Propulsion Laboratory, California Institute of Technology, Pasadena, CA 91109, USA}
 \email{daniel.p.cunnane@jpl.nasa.gov}

\date{\today}

\begin{abstract}
\noindent\hrulefill \\[0.5cm]

Thermal Kinetic Inductance Detectors (TKIDs) inherently combine the phonon-limited noise performance of traditional bolometers with the array scalability and responsivity of superconducting kinetic inductance detectors. Using a superconducting resonator as the thermally sensitive element provides high responsivity and tunable dynamic range, with phonon noise set by the cryogenic operating temperature of the free-standing membrane. In this work, MgB$_2$-based TKIDs are demonstrated operating from below 1 K up to 20 K  with characterized noise-equivalent power (NEP)  using integrated on-membrane heaters. A comprehensive characterization of electrical, thermal, and noise properties is presented. Phonon noise–limited performance is demonstrated from 4 to 8 K.

\end{abstract}
\maketitle

\onecolumngrid
\vspace{-3pt}
\hrule height 0.4pt
\vspace{5pt}
\twocolumngrid


Kinetic inductance–based thermal detectors have emerged as a promising technology for scalable, high-sensitivity imaging systems operating from millimeter-wave through far-infrared wavelengths. In particular, thermal kinetic inductance detectors (TKIDs) combine the intrinsic sensitivity of cryogenic bolometers with a microwave multiplexed readout architecture \cite{2UlbrichtTKID,5DAYArray,6DoyleKID}, offering a path toward large-format detector arrays for next-generation remote sensing and imaging applications \cite{AlTKID,3DANYBCO,4Timofeev_2014}.  

A major challenge in extending cryogenic detector technologies beyond low-background astrophysical instruments is the significant mismatch between ultra-sensitive low-temperature detectors and real-world imaging environments dominated by high optical backgrounds at or near 300 K. Traditional cryogenic detector platforms, including transition-edge sensors (TESs) and low-$T_c$ aluminum-based kinetic inductance detectors (KIDs), are optimized for photon-starved conditions such as cosmic microwave background (CMB) observations and deep-space astronomy \cite{7ECHo,11TES,21dekorte2003time,22irwin2002time,23dobbs2012frequency, 24abitbol2018ebex}. In Earth-observing and atmospheric imaging environments, the background photon noise limit in the mm-wave region is orders of magnitude higher, but there remains a gap in detector technology to get near or below this photon noise limit ($\sim fW/\sqrt{Hz}$) without sub-K systems.

Thermal detectors fabricated with novel superconducting materials offer an intermediate operating regime by enabling detector operation at high temperatures while maintaining background-limited sensitivity. This, therefore, reduces cryogenic requirements and improves compatibility for high-background imaging systems. This capability is particularly attractive for Earth observation, atmospheric monitoring, security screening, and suborbital and orbital remote sensing platforms operating across the millimeter-wave to far-infrared bands \cite{3DANYBCO}.

Recent advances in materials such as NbN and YBa$_2$Cu$_3$O$_7$ (YBCO)  have demonstrated kinetic inductance devices operating well above 4 K \cite{4Timofeev_2014,3DANYBCO}. Magnesium diboride (MgB$_2$) offers further advantages due to its relatively high critical temperature ($T_c$), large superconducting energy gap, and favorable microwave properties \cite{Kim2024,Greenfield2025, 25DANMGB2}.  High microwave quality factors in these films enable strong temperature responsivity in resonators while suppressing readout noise and achieving phonon-noise limited performance. The novel integration of MgB$_2$ inductors on free-standing silicon nitride (SiN$_x$) membranes enables thermal isolation suitable for TKID implementation, opening a temperature region for MgB$_2$-based devices to operate between ultra-low-background astrophysical instruments and room-temperature imaging systems.

A comprehensive performance characterization of MgB$_2$ TKIDs is presented in this paper. The prototype devices are designed conservatively to prioritize robustness and reproducibility over ultimate sensitivity. The bolometer thermal conductance, time constant, and inherent noise are characterized. The measured noise approaches the phonon noise limit over a broad temperature range, limited only by the designed coupling parameter rather than an intrinsic limitation. The noise measurements demonstrate the suitability of MgB$_2$ TKIDs for high-performance remote sensing and imaging applications.

\begin{figure}[t]
\centering
\includegraphics[width=0.85\columnwidth]{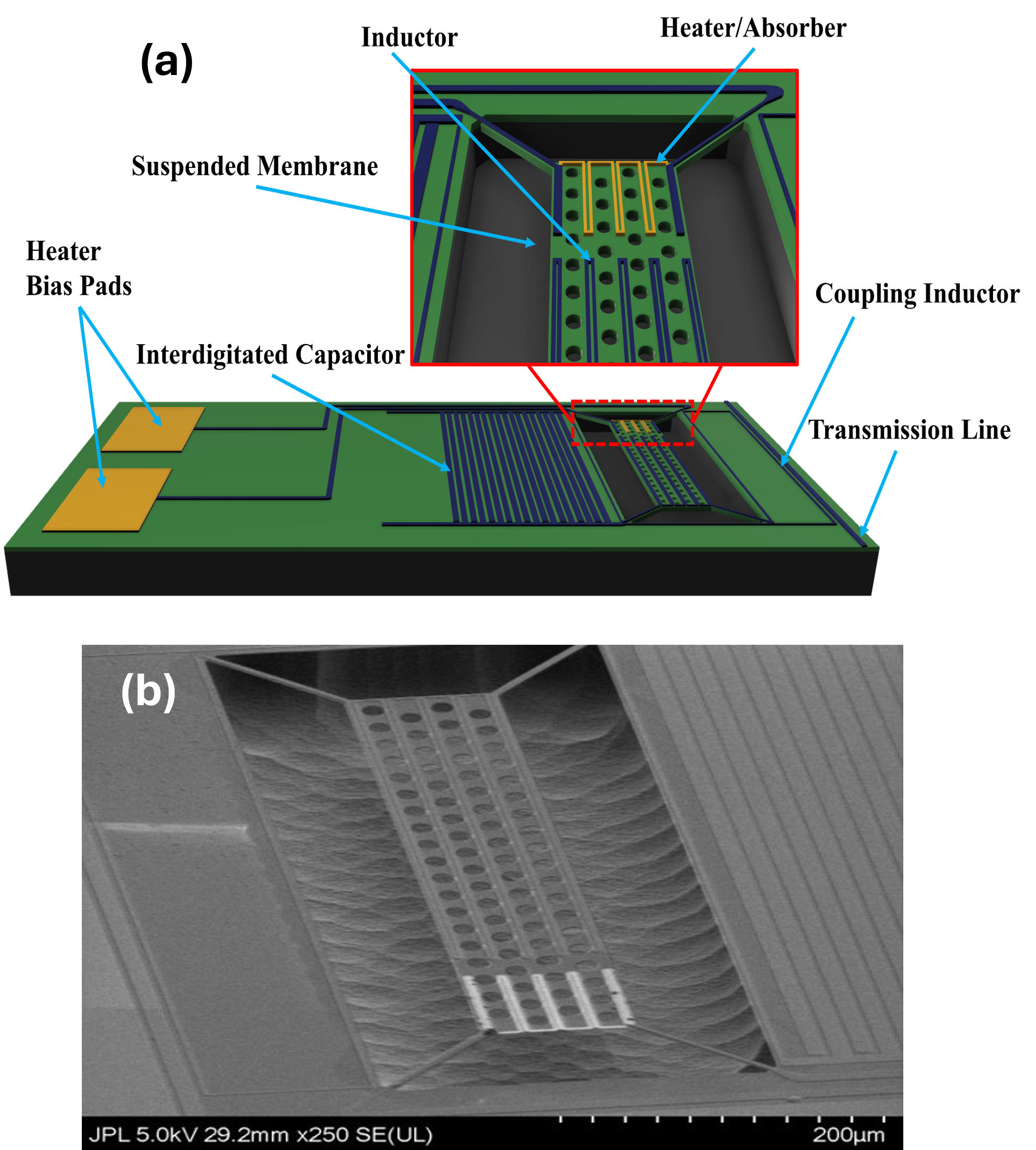}
\caption{ \justifying MgB$_2$ TKID design, (a) schematic diagram of the TKID, and (b) SEM of a released TKID membrane.}    
\label{fig:devdesign} 
\end{figure}

MgB$_2$ TKIDs are lumped-element resonators comprising of an interdigitated capacitor (IDC) and a discrete inductor coupled to a microwave feedline, as illustrated in Fig. \ref{fig:devdesign}. TKIDs exhibit a temperature-dependent resonant frequency given by $f_0(T) = 1/2\pi\sqrt{L(T)C}$, determined by the total capacitance of the IDC and the sum of the geometric inductance and superconducting kinetic inductance of the MgB$_2$ inductor, $L_{total} = L_{g} + L_{k}(T)$. The key feature of TKIDs is the suspension of the inductor on a thermally isolated membrane. Due to temperature dependence of $L_k$, heating of the membrane increases $L_k$ and shifts the resonant frequency downward. For the prototype devices, $L_k(T=4.2 \text{ K}) \approx 7.5 \text{ pH}/\square$, making the kinetic inductance fraction, $\alpha > 90\% $. The temperature dependent behavior enables TKIDs to operate as highly sensitive thermal detectors, converting absorbed power into measurable shifts in $f_0$. This architecture also allows for the multiplexing of as many as 2,000 TKIDs on a single readout line using microwave frequency-division multiplexing\cite{RFSOC2022}.

A five-device MgB$_2$ test chip is fabricated with varying IDC finger counts while maintaining identical inductors and membranes. The resulting capacitance variation produced resonant frequencies ranging from 850 MHz to 1.1 GHz at 4 K operating temperature, within 5\% of the designed frequencies and consistent with finite-element simulations. The membrane is attached to the main wafer via four legs, creating a weak thermal link between the membrane and the wafer. The thermal conductance, $G$, is determined by the cross-sectional area and the length of the legs. Although the leg geometry is designed with dimensions of 10 $\mu$m wide by 100 $\mu$m long, the membrane etch resulted in a final width of about 5 $\mu$m with minimal length variation.

Fabrication of the MgB$_2$ thermal detectors is designed to preserve film quality while minimizing oxidation throughout the process.  The devices are fabricated on 500 nm of ultra-low-stress SiN$_x$, forming the free-standing membrane  with a 100 nm thermal oxide layer beneath serving as an etch stop for the topside membrane etch, on an undoped high-resistivity Si wafer. The MgB$_2$ is deposited and annealed as previously reported \cite{Kim2024}. Following the annealing,  the Boron capping layer is etched, and  a hardmask layer of SiN$_x$ is deposited using plasma-enhanced chemical vapor deposition (PECVD).   The pattern is then transferred into the hardmask using a fluorine-based reactive-ion etch (RIE), followed by a BCl$_3$ plasma etch to transfer the pattern into the MgB$_2$ layer. The patterned MgB$_2$ is then passivated with an additional PECVD SiN$_x$ layer to protect the device during subsequent steps\cite{Greenfield2025}. This results in a SiN$_x$ thickness of about 750 nm, contributing to the thermal properties of the bolometer.

The on-membrane heaters and bond pads are patterned using a liftoff process where a negative resist pattern exposes the bond pad areas. The SiN$_x$ is then etched to expose the   MgB$_2$, followed by deposition of a 150 nm thick gold layer. The final liftoff step leaves the gold layer only in the patterned area. To accommodate fragile suspended membranes, slots on the backside of the wafer are preemptively patterned and etched to enable manual separation of individual chips after fabrication. Holes are patterned over the membrane area to ensure a homogeneous release, and a fluorine-based etch is used to penetrate through the SiN$_x$/SiO$_2$ and expose the underlying silicon. The membrane is released by etching the underlying silicon substrate with XeF$_2$, which selectively removes the silicon and stops in the 100 nm oxide layer. A scanning electron micrograph (SEM) of a membrane released from substrate with the MgB$_2$ inductor and gold heater resistor patterned onto the membrane is shown in Fig. \ref{fig:devdesign}(b).


MgB$_2$ TKIDs are placed in a custom copper package and characterized in a closed cycle cryo-cooler. Measurements are performed at NASA Jet Propulsion Laboratory (JPL) using a He$^4$ sorption refrigerator to achieve temperatures as low as 750 mK and at Arizona State University (ASU) using a He$^3$ sorption refrigerator to achieve temperatures as low as 250 mK. {Vector} network analyzers (VNA) for resonator characterization are used at both labs. At JPL, a standard homodyne setup is used to measure the noise and time constants in single resonators without correlated noise removal. At ASU, a Radio Frequency System-on-Chip (RFSoC) readout system \cite{RFSOC2022} is used  to measure the time constant and intrinsic detector noise. The TKID's response is measured using ten microwave tones: one on-resonance and nine off-resonance, used to construct a correlated noise template. The on-resonance response is then projected onto this template, and the result is subtracted to reduce noise common to the ten tones.

\begin{figure}[t]
\includegraphics[width=1\columnwidth]{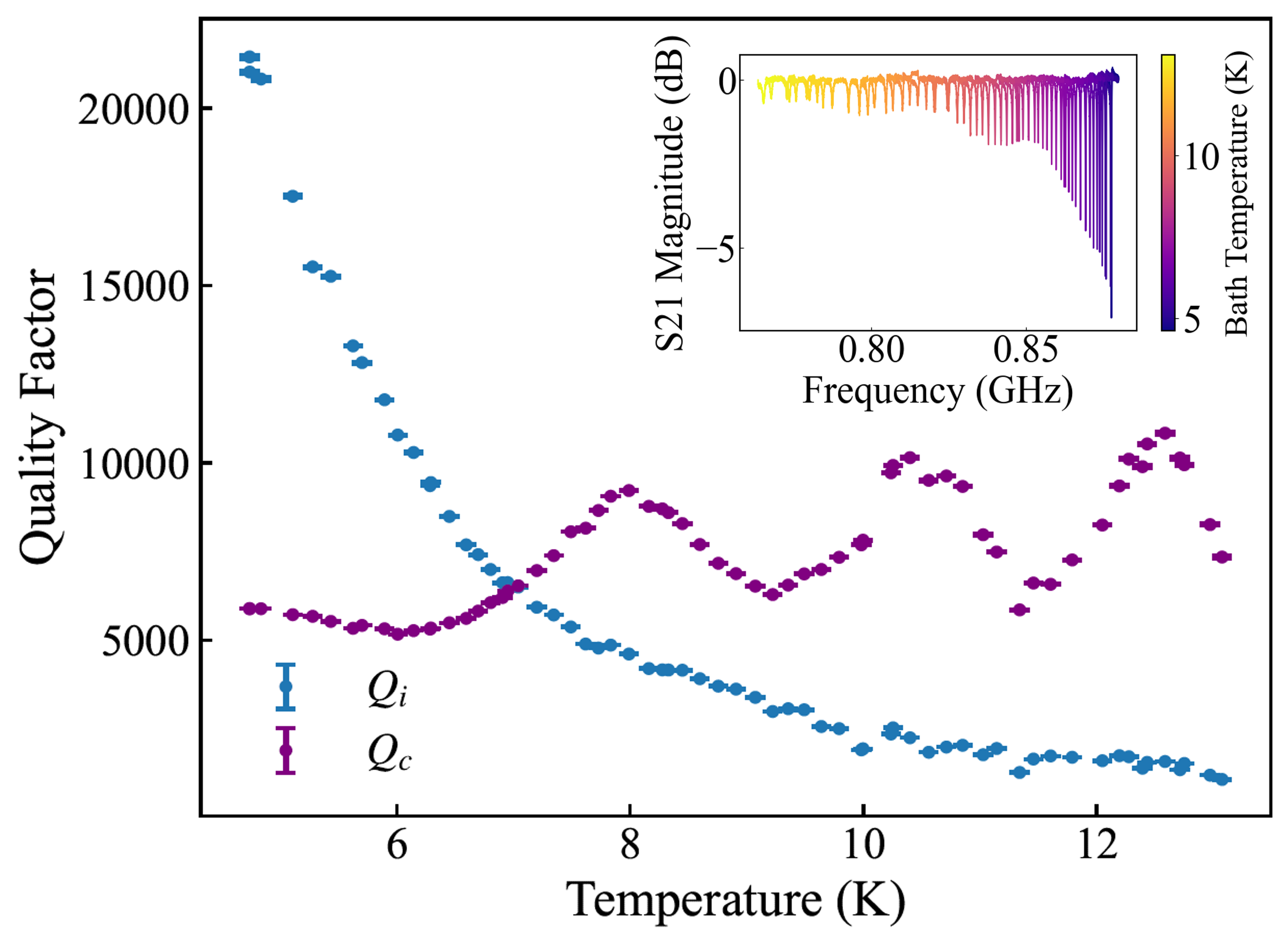} 
\caption{  \justifying Measured and fitted  quality factors of an MgB$_2$ TKID resonator at different temperatures. Inset shows the $S_{21}$ magnitude vs frequency over the same temperature range for the  device.}  
 \label{fig:qfactors} 
 \end{figure}

The internal and coupling quality factors of $Q_i>2\times10^4$ at 4.2 K bath temperature are shown in Fig. \ref{fig:qfactors}, with negligible degradation from membrane release. The temperature sensitivity is maximized when the coupling quality factor ($Q_c$) matches the internal quality factor ($Q_i$), corresponding to a 6 dB deep resonance. $Q_c$ is related to the design of the resonator and coupling to the transmission line rather than temperature-dependent effects. At low temperatures the total quality factor, $Q_r$, is dominated by $Q_c$ and gradually decreases as $Q_i$ is reduced. 
The $S_{21}$ transmission of an MgB$_2$ TKID for a wide range of operating temperatures is shown in Fig. \ref{fig:qfactors} (inset), highlighting the strong temperature dependence of the resonance. The resonator quality factor directly impacts detector responsivity. Although achieving phonon noise below the fW/$\sqrt{\mathrm{Hz}}$ range typically requires high $Q$, noise as low as 6 fW/$\sqrt{\mathrm{Hz}}$ is demonstrated in NbN devices with quality factors as low as 100 near the $T_c$ of the superconductor, where low $Q$ is offset by the strong temperature dependence of $L_k$ \cite{4Timofeev_2014}.

The net thermal power flowing from the bolometer membrane through the supporting legs to the surrounding wafer, which acts as a thermal bath, is modeled as $P_{\text{leg}} = K(T^n - T^n_{\text{bath}})$, where $T$ is the membrane temperature, $T_{\text{bath}}$ is the temperature of the thermal bath, and $K$ is a material and geometry dependent constant\cite{Mather:82}. The thermal conductance is given by the derivative with respect to the membrane temperature,  $G(T) = \frac{dP_{\text{leg}}}{dT} = nKT^{n-1}$.  
Power $P_H$ is applied to the bolometer by applying a known voltage to the on-membrane heater, and $P_{\text{leg}} = P_H$ when the membrane reaches an equilibrium temperature. The cold resistance of the heater $R_H$ is measured at ASU to be $R_H =$  11.200 $\pm$ 0.005 $\Omega$, using a four-point resistance measurement. For a given applied power, the membrane temperature $T$ is then determined by comparing the resonant frequency to shifts associated with only changes in the bath temperature $f_0(T_{\text{bath}})$, see e.g. the inset of Fig. \ref{fig:qfactors}. 

\begin{figure}[t]
\includegraphics[width=1\columnwidth]{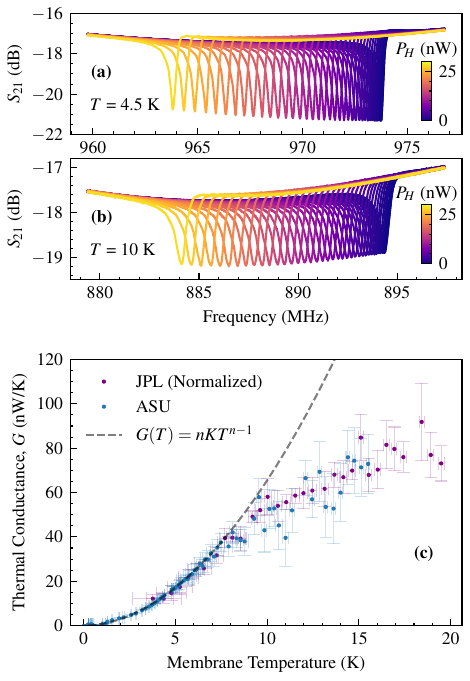} 
\caption{ \justifying  TKID performance as a function of heater power $P_H$ from 250 mK to 20 K.  TKID $S_{21}$ sweeps with different DC heater powers applied at a bath temperature of (a) 4.5 K  and  (b) 10 K. These measurements are used to calculate $G$ for a specific bath temperature. (c) Measured thermal conductance of the TKID. The heater resistance for the TKID at JPL is normalized to match the thermal conductance of the device measured at ASU. Both measurements exhibit nearly identical temperature dependence.}
 \label{fig:thermal_conductance}   
\end{figure}

At each bath temperature, $S_{21}(f)$ sweeps centered on the TKID resonant frequency are recorded with a VNA while DC voltages are applied, corresponding to dissipated heater powers $P_H$  from 0 to 28 nW. Increasing $P_H$ shifts the resonance to lower $f_0$, as shown in Fig. \ref{fig:thermal_conductance}(a) and Fig. \ref{fig:thermal_conductance}(b) at 4.5 and 10 K, respectively. Since 60\% of the inductor is located on the prototype TKID membrane, the frequency shift arises from only this fraction of the inductor heated by $P_H$. The membrane temperature changes can then be accurately approximated by using the bath temperature dependence of the resonator and adjusting for the fraction of the inductor on the membrane, $\frac{d f_0}{dT} \approx F_{\text{m}}\frac{d f_0}{d T_{\text{bath}}}$, where $F_{\text{m}} = 0.6$ is the inductor fraction on the membrane. For each bath temperature, the thermal conductivity $G(T)$, shown in Fig. \ref{fig:thermal_conductance}(c), is  determined for small heater power  by
\begin{equation}\label{eq:1}
G(T) = \frac{d P_H}{d f_0} \frac{d f_0}{d T} = F_{\text{m}} \frac{d P_H}{d f_0} \frac{d f_0}{d T_{\text{bath}}}.
\end{equation}

From 250 mK to 7 K, $G \propto T^{1.93}$ and closely follows $G(T) = nKT^{n-1}$ for $n = 2.93 \pm 0.03$ and $K = 0.260 \pm 0.013$ nW/$\mathrm{K}^{n}$. This behavior does not hold for higher temperatures, which has been reported previously\cite{holmes1998measurements,hoevers2005radiative,Zink2004}. The slowed increase of the TKID $G(T)$ at higher temperatures reflects a transition from ballistic to diffusive phonon transport in the support legs,  limiting heat transfer across the membrane.

\begin{figure}[t]
\includegraphics[width=1\columnwidth]{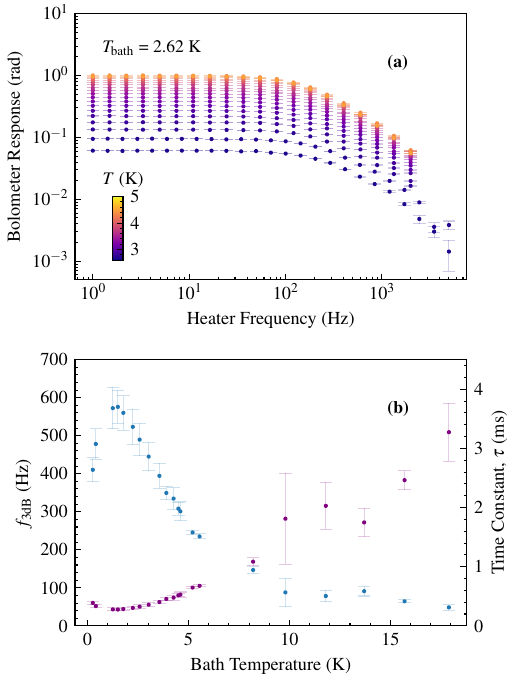} 
\caption{ \justifying Thermal response   as a function of membrane temperature, (a) TKID response vs. membrane temperature at a fixed bath temperature under different DC heater powers, and  (b) response rolloff frequency (left axis) and time constant (right axis) measured at fixed bath temperatures from 250 mK to 17.9K.}
 \label{fig:time_constant} 
\end{figure}

The bolometer time constant, $\tau_ {\mathrm{bolo}} = C(T)/G(T)$, determines the   detector bandwidth. The device response follows a single-pole low-pass transfer function with a rolloff frequency $f_{3\mathrm{dB}}$, consistent with a lumped-element thermal model, given by $f_{3\mathrm{dB}} = 1/(2\pi\tau_{\mathrm{bolo}})$. TKID responsivity rolloff at a bath temperature of 2.62 K is shown in Fig. \ref{fig:time_constant}(a).  
Sinusoidal heater modulation from 1 Hz to 1 kHz is applied to measure responsivity as a function of frequency. The rolloff frequency $f_{3\mathrm{dB}}$ is determined by fitting the response to a normalized first-order lorentzian. Using a DC offset in the applied heater power enables measurements at different effective membrane temperatures, $T$, indicating an increase in  the time constant  at higher temperatures. $\tau_{\mathrm{bolo}}$ deviates from measurements obtained by changing $T_{\mathrm{bath}}$ due to temperature gradients across the legs.

The measured time constant exhibits a  temperature dependence, increasing from approximately 0.3 ms at temperatures below 1 K to approximately 3  ms at 18 K.  This corresponds to a rolloff frequency that begins near 570 Hz at low temperatures and gradually decreases to 50 Hz as the membrane temperature increases, shown in Fig. \ref{fig:time_constant}(b). 
The non-monotonic changes in $\tau_ {\mathrm{bolo}}$ with temperature reflects the combined temperature dependence of the membrane heat capacity and thermal conductance. At temperatures below 0.5 K, $\tau_ {\mathrm{bolo}}$ increases, consistent with heat capacity changes associated with two-level systems \cite{TLS_SiNx}. Despite this increase, the measured time constants remain well within the bandwidth requirements of millimeter-wave and far-infrared remote sensing instruments.


    %


The noise-equivalent power (NEP) characterizes the sensitivity of the MgB$_2$ kinetic inductance bolometers and is compared to the theoretical phonon noise. Photon noise introduces fluctuations in the number of incident photons, leading to measurement uncertainty, particularly in single-mode detectors. In thermal detectors, additional fluctuations arise from phonon-mediated energy transport within the bolometer material. This phonon noise arises from stochastic thermal energy exchange between the bolometer membrane and the thermal bath. The NEP contribution from phonon noise \cite{Mather:82} is 
\begin{equation}\label{eq:2}
    NEP_{ph}^2 = 4F(T,T_{\text{bath}})k_B T^2 G(T),
\end{equation}
where  F$(T,T_{\text{bath}})$ is a factor that accounts for the temperature gradient between the bolometer and the thermal bath, $k_B$ is Boltzmann's constant, $T$ is the operating membrane temperature of the bolometer, and $G(T)$ is the thermal conductance at operating temperature.   
Controlling phonon noise is crucial for maximizing the sensitivity of TKID bolometers. Phonon noise can be minimized by designing the thermal link with  lower conductance or operating the bolometer at lower temperatures.

\begin{figure}[t]
    \centering
\includegraphics[width=1\columnwidth]{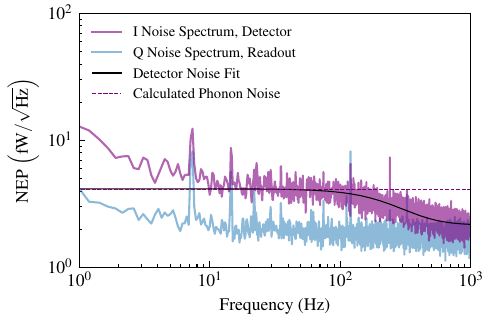}
    \caption{  \justifying NEP spectrum of the TKID noise at 4.56 K. The in-phase (I) and quadrature (Q) components of the noise signal are detector and readout noise, respectively.}    
   \label{fig:noise_spectrum}
\end{figure}

The NEP spectrum of the TKID at 4.56 K with no heater power applied to the heater is shown in Fig. \ref{fig:noise_spectrum}. The detector noise is fitted with a single pole, low-pass transfer function added in quadrature to a constant noise term. From the fit, the rolloff frequency is $f_{3\mathrm{dB}} = 323.3 \pm 4.7$ Hz and the measured detector NEP from 10 Hz to 50 Hz is 4.17 $\pm$ 0.04 fW/$\sqrt{\mathrm{Hz}}$. The rolloff frequency is in close agreement with the results in Fig. \ref{fig:time_constant}(b).  
The measured noise is consistent with the expected phonon noise of 4.1 fW/$\sqrt{\mathrm{Hz}}$, calculated from the measured $G(T)$ using Eq.~\eqref{eq:2} with $T = T_{\text{bath}}$ and $F(T,T_{\text{bath}})=1$.  Above the rolloff frequency, the detector noise spectrum approaches a constant effective noise power of 2.00 $\pm$ 0.03 fW/$\sqrt{\mathrm{Hz}}$, shown in Fig. \ref{fig:noise_spectrum}, corresponding to the equivalent contribution from readout noise.


\begin{figure}[t]
\includegraphics[width=1\columnwidth]{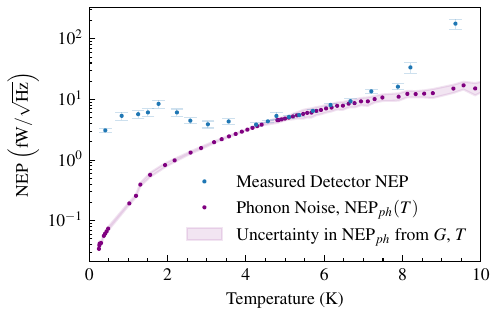} 
\caption{  \justifying Detector NEP with no heater power and temperature of $T = T_{\text{bath}}$. }
 \label{fig:NEP} 
\end{figure}

The measured detector NEP as a function of membrane temperature from the millikelvin regime up to 9.8 K is shown in Fig. \ref{fig:NEP}. The detector NEP at each temperature is obtained by median-filtering the NEP spectrum and computing the mean and standard deviation over 10 Hz to 50 Hz. At low temperatures, the detector NEP deviates from the expected phonon noise limit due to the low temperature dependence of the resonant frequency below 4 K and limitation of $Q$ by $Q_c$. Phonon noise–limited operation is achieved over the technologically relevant 4 K to 8 K range. At higher temperatures, the increase in detector NEP reflects the combined temperature dependence of the membrane heat capacity and the thermal conductance of the support legs. Despite an increasing phonon noise limit at high operating temperatures, the degradation of $Q_i$ leads to undercoupled ($Q_c>Q_i$) shallow resonators with smaller responsivity. $Q_c$ can therefore be redesigned for high responsivity across the operating temperature of the resonator. 

The   development and characterization of magnesium diboride thermal kinetic inductance detectors operating over an extended temperature range from below 1 K up to 20 K are presented. Phonon noise–limited performance is demonstrated over the technologically relevant 4 K to 8 K, with $\mathrm{NEP}_{\mathrm{measured}} < 10^{-14}$ W/$\sqrt{\mathrm{Hz}}$ at $T < 8 \text{ K}$. 
The bolometer time constant of $\tau \approx 0.7$ ms at 5.6 K indicates that the thermal conductance can be further reduced to achieve lower NEP while maintaining sufficient bandwidth for practical imaging applications. MgB$_2$ TKIDs represent a promising detector technology for millimeter-wave to far-infrared imaging systems by enabling elevated operating temperatures, scalable multiplexed readout, and phonon-limited sensitivity. The intrinsically frequency-agnostic architecture  enables compatibility with a wide range of optical coupling schemes, supporting array-level integration and system-level implementation.
\vspace{2mm}

This research was carried out at the Jet Propulsion Laboratory, California Institute of Technology, under a contract with the National Aeronautics and Space Administration (80NM0018D004). The work in this publication was supported by NASA’s Astrophysics Research and Analysis Program (APRA) under task order No. 80NM0018F0610, and proposal number 20-APRA20-0127. This work was made possible by the capabilities developed by Dr.\ Daniel Cunnane under a Nancy Grace Roman Early Career Technology Fellowship (RTF). Dr.\ Tahereh Jabbari's research was supported by appointment to the NASA Postdoctoral Program at the Jet Propulsion Laboratory, administered by Oak Ridge Associated Universities under contract with NASA. We acknowledge the support and infrastructure provided for this work by the Microdevices Laboratory at JPL, as well as the funding provided by the JPL SIP. The authors would like to thank Warren Holmes, Ritoban Basu Thakur, Rick LeDuc, Peter Day, Jack Sayers and Bruce Bumble for helpful discussions along the way. © 2026. All rights reserved.

\nocite {*}
\bibliographystyle{ieeetr}
\bibliography{Ref}

\end{document}